\documentclass[12pt,twoside]{article}
\usepackage{cmp2e,epsfig,graphics}

\title[Public transport networks
]{Scaling in public transport networks
}

\author[C. von Ferber, Yu. Holovatch, V. Palchykov]{
  C. von Ferber\refaddr{Freiburg},
  Yu. Holovatch\refaddr{ICMP,Linz,LUniv},
  V. Palchykov\refaddr{LUniv}
 }

\addresses{
  \addr{Freiburg} Theoretische Polymerphysik, Universit\"at
                  Freiburg, D-79104 Freiburg, Germany
  \addr{ICMP}  Institute for Condensed Matter Physics,
               National Acad. Sci. of Ukraine,\\
               UA-79011 Lviv, Ukraine
    \addr{Linz}  Institut f\"ur Theoretische Physik,
               Johannes Kepler Universit\"at Linz,\\
               A-4040 Linz, Austria
  \addr{LUniv} Ivan Franko National University of Lviv,
               UA-79005 Lviv, Ukraine
}
\begin{document}

\maketitle

\begin{abstract}
We analyse the statistical properties of public transport networks.
These networks are defined by a set of public transport routes (bus
lines) and the stations serviced by these.  For larger networks these
appear to possess a scale-free structure, as it is demonstrated
e.g. by the Zipf law distribution of the number of routes servicing a
given station or for the distribution of the number of stations which
can be visited from the chosen one without changing the means of
transport.  Moreover, a rather particular feature of the public
transport network is that many routes service common subsets of
stations. We discuss the possibility of new scaling laws that govern
intrinsic features of such subsets.
\pacs 89.75.-k, 89.75.Da, 05.65.+b
\end{abstract}
\section{Introduction. What are complex networks for a physicist \label{I}}

Complex networks have only recently become a subject discussed on
the pages of physical journals \cite{books}. However, currently
the statistical mechanics of complex networks is an important and
quickly evolving field of physics \cite{Barabasi02}, as one can
check e.g. making a search in the WWW (the latter being another
huge complex network and hence by itself an important subject of
study).  As has been worked out in the meantime, complex web-like
structures are involved in such different systems as the already
mentioned WWW (with its documents as nodes and links as edges)
\cite{www}, the metabolism of a biological cell (substrates
connected by bio-chemical reactions) \cite{cell}, social
communications (human beings connected by various relationships)
\cite{social}, ecological systems (food webs joining different
species) \cite{ecol}, etc. Therefore, these and similar systems
can be formally described in terms of the same formalism and very
often they manifest similar statistical behaviour.

Of particular interest for our study is the question if the network
displays scale-free properties: a notion introduced to characterize
a network which does not posses a typical scale \cite{Barabasi99}. A
network is called scale-free when its node degree distribution,
i.e. the probability that a randomly selected node has $k$ edges, possesses
a power-law tail:
\begin{equation} \label{1}
P(k) \sim k^{-\gamma}.
\end{equation}

Since its first observation by G.K. Zipf in quantitative linguistics
\cite{Zipf35} the power law (\ref{1}) is referred to as Zipfs law. For a
physicist working in the field of statistical physics, appearance of
universal power laws - scaling laws - serves as a manifestation of
collective behaviour of a many-particle system at the critical point
\cite{critbooks}. This explains, in particular, why physicists
(often those, working in a field of critical phenomena) apply their efforts
and skills in the network theory and why these efforts may be fruitful.

The last decades of the past century offered theoretical descriptions
of critical phenomena which show why universal scaling laws (\ref{1})
emerge and give reliable numerical estimates for the exponents
$\gamma$ governing the scaling of different physical observables
\cite{order}. This description was made possible by the application of
field theoretical methods in many particle physics \cite{rgbooks} and
serves as a background to explain and to predict scaling in various
systems. However, the modern theory of networks differs from the
modern theory of critical phenomena in a way that although it operates
with models describing different types of networks and looks for the
Zipf laws governing the scaling of the properties of these networks, the prediction or
explanation of certain scaling properties is done mainly via
computer simulations, or a simple (mean-field like) analysis. A
theoretical description of complex networks, involving e.g. a theory
beyond the tree graph approximation similar to the field-theoretical
description of critical phenomena with non-trivial interactions
\cite{critbooks,order,rgbooks} is still missing.

At this level of network analysis it is important both to search
for new types of networks that exist in complex structures as well
as to collect ``empirical data": to look for observables
describing these networks and to analyse their properties. In our
paper, we want to attract attention to a feature frequently
encountered in complex networks: Looking at the paths of
connections on the ``motherboard'' of a computer, the wiring in a
car, or even the neural connections along the spine of vertebrates
one observes as a common feature that the physical paths used by
the lines connecting different nodes are often shared by many
other lines connecting other nodes. We are interested in the
distribution of the load along these paths. We study this property
for more easily accessible examples of networks of public
transport (PT networks). We will show that a PT  network may
demonstrate a scale-free behaviour. Moreover, a rather particular
feature of the PT network is that many routes possess common
subsets of stations. We will demonstrate that new scaling laws may
govern intrinsic features of distributions defined on these
subsets.

The rest of the paper is organized as follows: in the next section \ref{II}
we explain what we mean by a PT network, introduce the observables
used to describe it and give the sources of our further analysis. In section
\ref{III}
we will show that the node degree distribution of PT networks obeys the
Zipf law
(\ref{1}) and hence a PT network may constitute an example of a scale-free
network \cite{Barabasi99}. In section \ref{IV} we will consider
in particular the traffic load distribution of paths of different length
studying situations
where many routes possess common subsets of
nodes. We check these properties for scale-free behaviour and
speculate on a generalization.
 Conclusions and an outlook are given in section \ref{V}.

\section{A PT network \label{II}}

In our study we consider networks of public transport
(buses, trams, and subways), called hereafter  PT networks.
In a PT network, the nodes are the stations of public
transport and the edges are the links connecting them along
the route (see figure \ref{fig1}). We will be interested in
various characteristics of PT networks that describe statistical
properties of node-link distributions. The examples are given below.
We will perform our study according to the following scheme:
\begin{figure} 
\begin{center}
\epsfxsize=3cm
\epsfysize=3cm
\epsffile{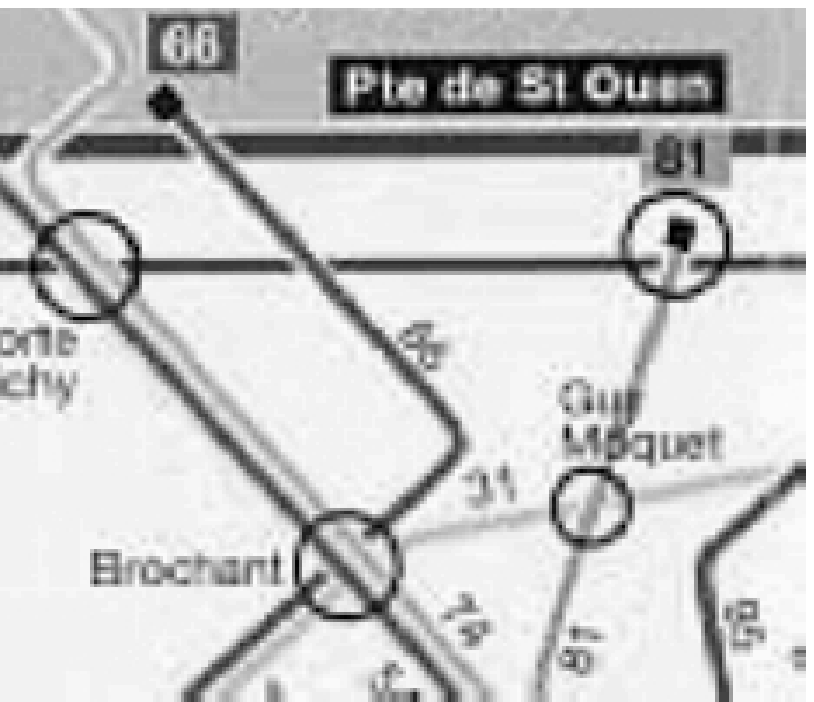}
\epsfxsize=3cm
\epsfysize=3cm
\hspace{3cm}
\epsffile{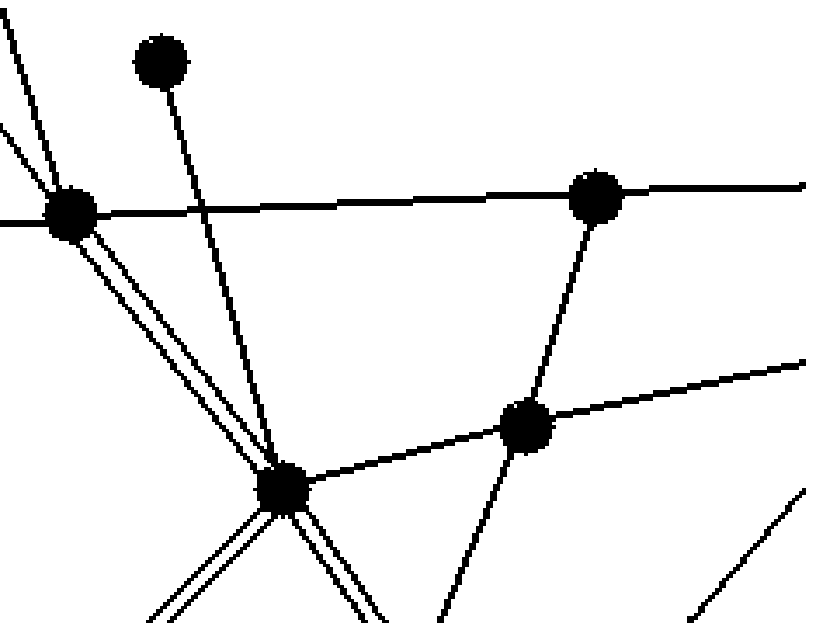}
\end{center}
\caption{\label{fig1} A part of the public transport scheme
for Paris   and a  PT network, which corresponds
to it.}
\end{figure}

\noindent
{\bf a)} choose a PT network;

\noindent
{\bf b)} make an ordered list of stations visited by each line;

\noindent
{\bf c)} check the network statistics.

\noindent
Let us comment on each of the above items before giving results
about the network structure.

{\bf a)}
As usual, to allow general properties of a network structure
to manifest themselves, the network analysed should be
large enough in terms of numbers of nodes and edges. Therefore, we choose
the PT networks of big cities,
having large numbers of routes and stops.
The results presented below are based on an analysis of PT
networks of Berlin (198 routes and 2952 stations), D\"usseldorf
(124 and 1615) and Paris (232 and 4003). In an extension
of our present study which is under way we analyse the PT networks
of more and larger cities.

{\bf b)}
The schedules of public transport for the above cities were
downloaded from the internet \cite{data} and brought into appropriate
format to construct the ordered lists
of stations serviced by each line
(hereafter we do not make any difference between bus/tram/subway
PT lines). These serve as a background for the network structure
analysis.

{\bf c)}
The ordered lists of stations allow to perform
some statistics checking typical quantities describing the network.
The most simple one concerns the number
of lines that service a given station \cite{note1}
and the number of other stations
one may reach without changing the bus/tram from a given station.
The distribution of the first quantity gives the familiar node degree
distribution $P(k)$ introduced in the previous section.
The choice of the second quantity comes about as a node degree
distribution of a conjugated network where each station (node) is connected
with all other stations for which there is a route servicing both.
This has quite practical consequences: it describes the neighbourhood of given
station and hence its "utility". Below, we will denote this
the size of this neighbourhood by $Z_1$.

A lot of different characteristics of the network can be introduced
depending on the particular problem one is interested in. Here, we want
to attract attention to a particular feature of a PT network: often,
a sequence of nodes is joined by more than one line.
This is the familiar situation when one can go from one station to
another by different train or bus lines without making a change. To study
the distribution of such sequences of stations, let us introduce the quantity
$P(L,N)$: the number of node segments of length $L$ connected by
$N$ lines.

Results of numerical analysis of the above introduced quantities
$P(k)$ and $P(L,N)$ will be presented in the next two sections.
\begin{figure} 
\begin{center}
\epsfxsize=7cm
\epsfysize=7cm
\epsfclipon
\epsffile{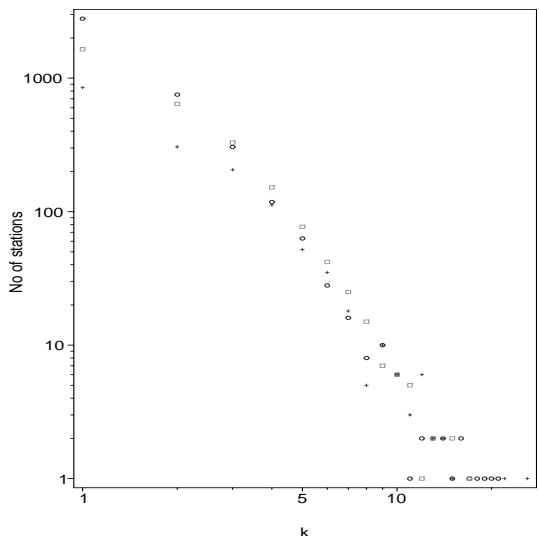}
\end{center}
\caption{\label{fig2} Number of stations as a function of the node degree
$k $ for the PT networks of Berlin (squares), D\"usseldorf (crosses),
Paris (circles). Being normalized, this function gives the node degree
distribution $P(k)$ \protect(\ref{1}).
}
\end{figure}
\section{Scale-free behaviour \label{III}}

First, we examine the node degree distribution $P(k)$ of the PT
networks. Results are shown in the figure \ref{fig2}, where we plot
the number of stations  for the PT networks of Berlin, D\"usseldorf,
and Paris as a function of the routes going through them (a node
degree $k$ \cite{note1}). Being normalized by the overall number of stations, this
function gives the node degree distribution $P(k)$,
hence both quantities obey the same scaling. One observes a power-law
behaviour (\ref{1}) in figure \ref{fig2} which leads to
the conclusion that the PT networks may be scale-free.
The least squares fit for
all data points of each city gives:
Berlin: 2.90; D\"usseldorf: 2.45; Paris: 2.94.
The values of the two larger networks are close to $\gamma=3$ corresponding
to the preferential attachment scenario \cite{Barabasi02}.

\begin{figure}[htbp]
\begin{center}
\includegraphics[width=72mm]{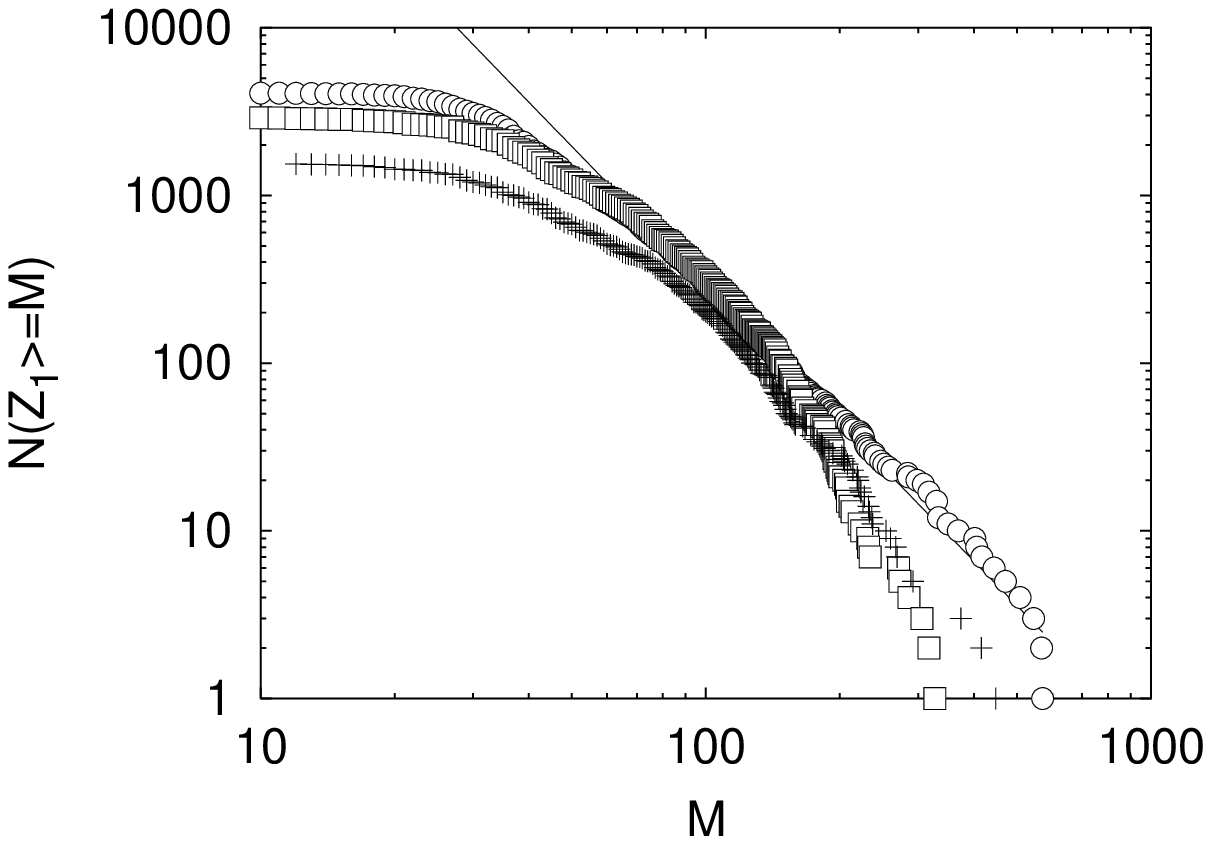}
\includegraphics[width=72mm]{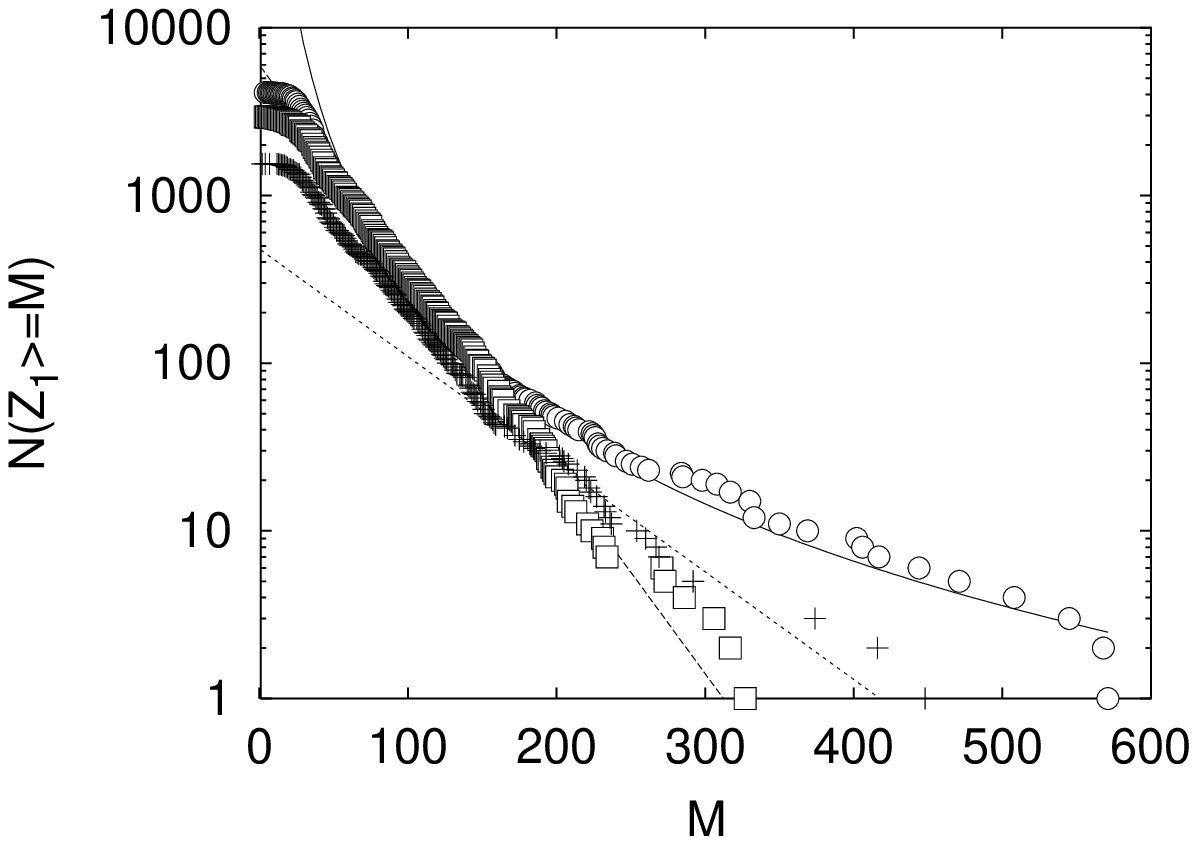}
\end{center}
\caption{\label{fig3} Integrated distribution of the direct
neighbourhood $Z_1$ of a given station.  Number $N$ of stations
with $Z_1\geq M$ for PT networks of Berlin (squares), D\"usseldorf
(crosses), Paris (circles). Left: double-logarithmic, right:
log-linear plot. }
\end{figure}

Another network property analysed here is the size $Z_1$ of the direct
neighbourhood
of a given station introduced in the
section \ref{II}: the number of stations one may reach without changing
from that station. By definition, for a given station this quantity is
obtained by counting all stations $Z_1$ which belong to the routes crossing it.
We show the number $N(Z_1\geq M)$ of stations with a neighbourhood larger than
$M$ as function of $M$ in figure \ref{fig3}. This function corresponding to
an integrated distribution is by definition
monotonous and thus smoother than the distribution of $Z_1$ itself.
As is seen from the double logarithmic and log-linear
plots only the largest (Paris) network develops a clear power law tail
with an exponent for the integrated distribution of about $\gamma-1=2.7$.
The Duesseldorf network may also be approximated by two exponentials.

Recently, another  PT network property was reported to
possess universal scaling behaviour: it was shown that the mean distance
between nodes of different degrees is governed by a scaling law \cite{Holyst04}.

\section{Segment distributions \label{IV}}

As we have already  seen  in the previous section, large PT
networks may have scale-free properties as shown by the power-law
behaviour of their node-degree distribution. Our next step will be
to study some other characteristics of the PT networks and to
check them for a power-law behaviour. Another example was given in
the previous section by the neighbourhood size $Z_1$. Here, we
will continue the analysis for the values $P(L,N)$, as introduced
in section \ref{II}: a number of node segments of length $L$
connected by $N$ lines. Our interest in these values is caused by
the fact, that for a real ``physical'' network different numbers
of links between nodes correspond to different loads on the
connection between these nodes. Besides the PT network an example
may be given by the network of wires connecting complex electric
circuit, or a network of tubes transmitting a fluid etc. In these
cases, the information about the loads on the links and their
distribution is important not only for understanding the entire
network features, but also for optimizing its structure.

\begin{figure}
\begin{picture}(200,550)
\put(10,401) {\includegraphics[width=6cm,height=6cm]{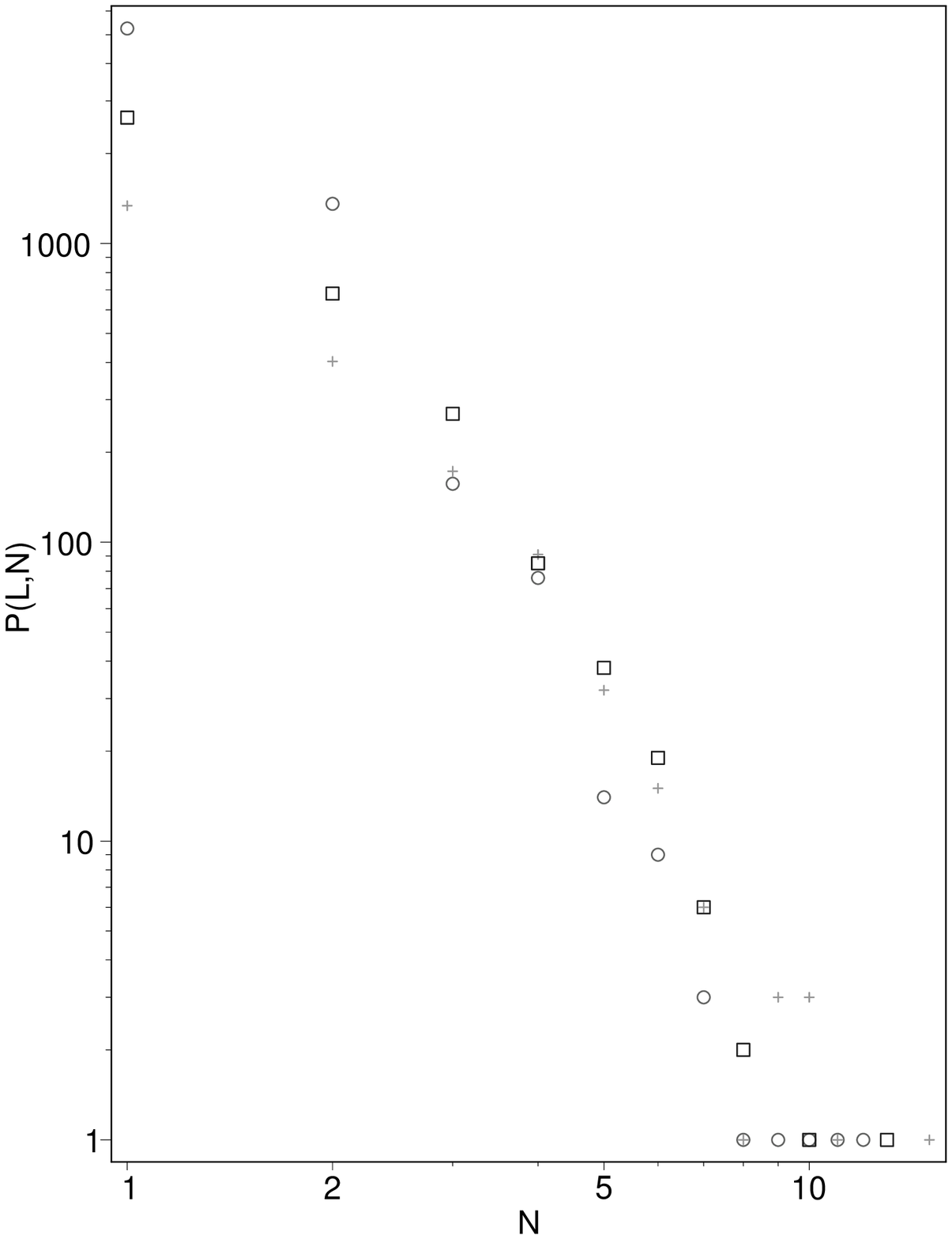}}
\put(75,386) {\parbox[t]{5mm}{\bf a.}} \put(215,401)
{\includegraphics[width=6cm,height=6cm]{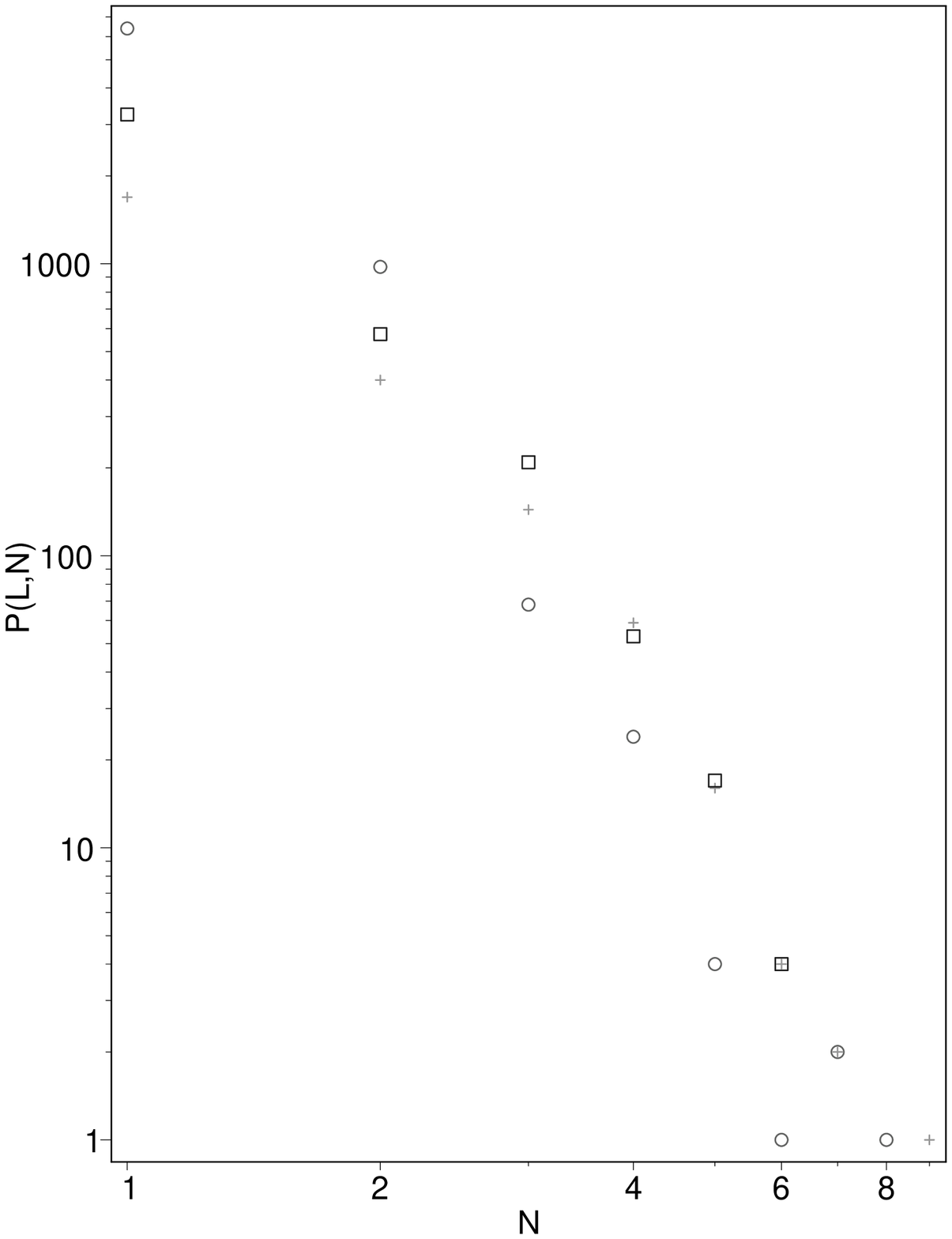}} \put(255,386)
{\parbox[t]{5mm}{\bf b.}} \put(10,210)
{\includegraphics[width=6cm,height=6cm]{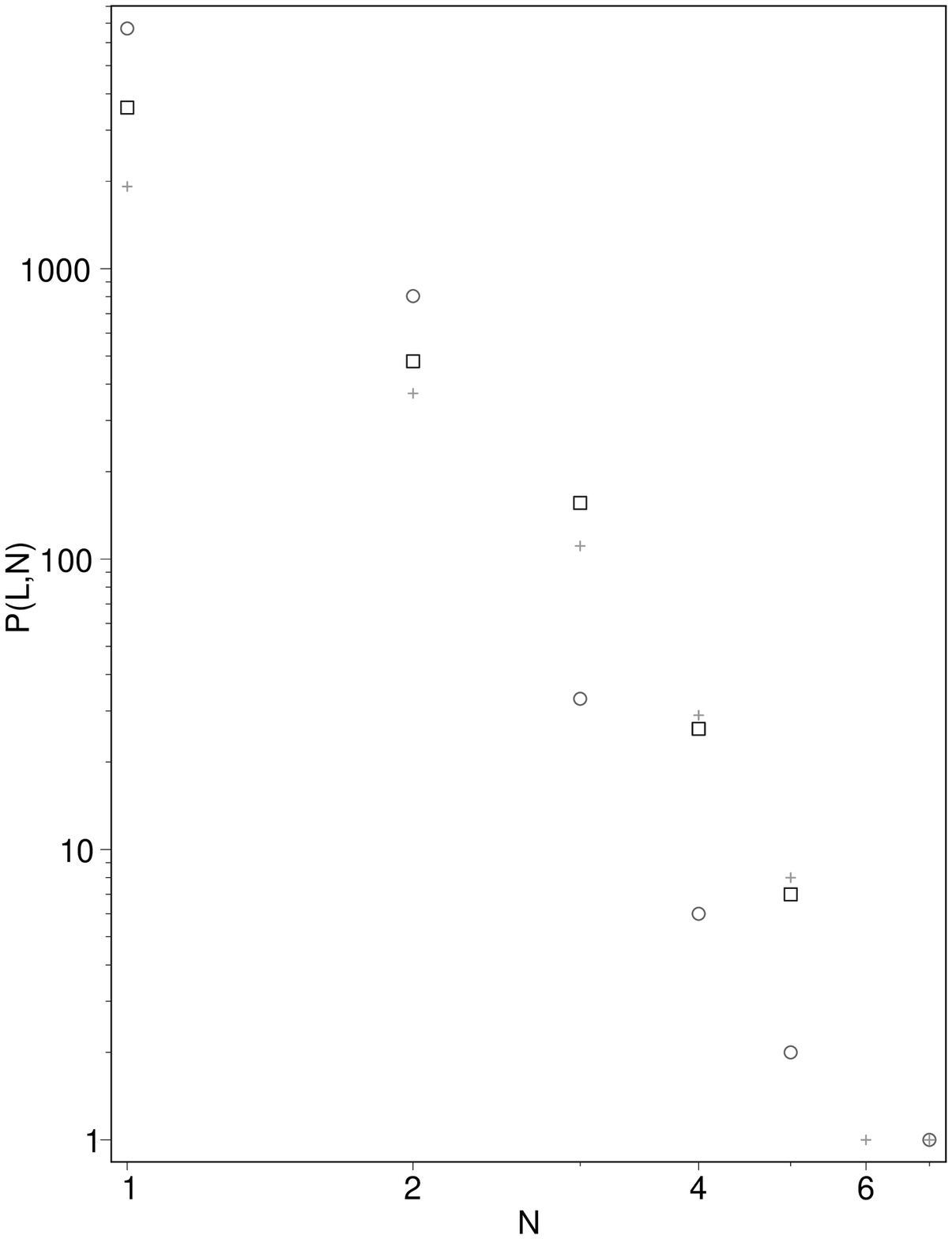}} \put(75,200)
{\parbox[t]{5mm}{\bf c.}} \put(215,210)
{\includegraphics[width=6cm,height=6cm]{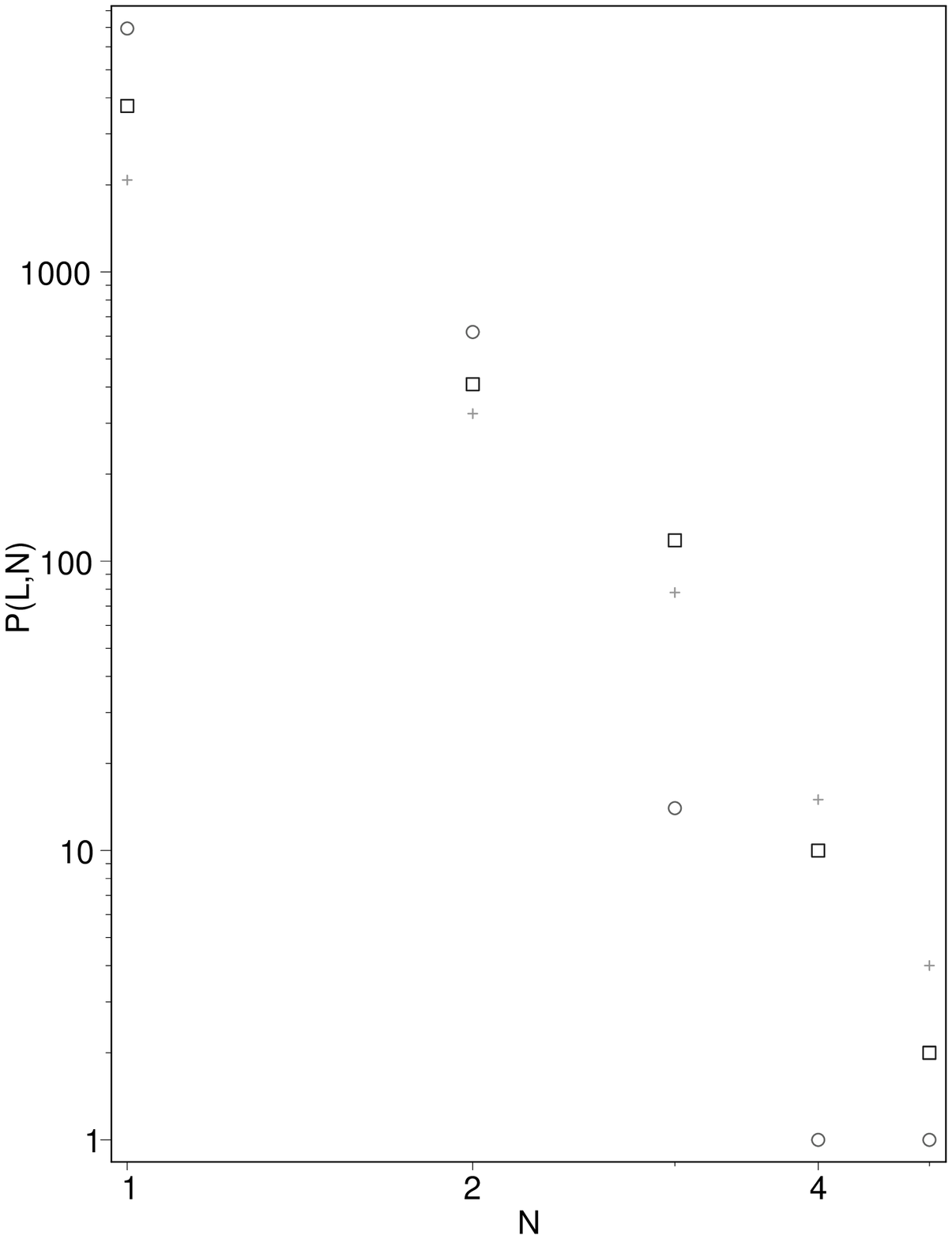}} \put(255,200)
{\parbox[t]{5mm}{\bf d.}} \put(10,10)
{\includegraphics[width=6cm,height=6cm]{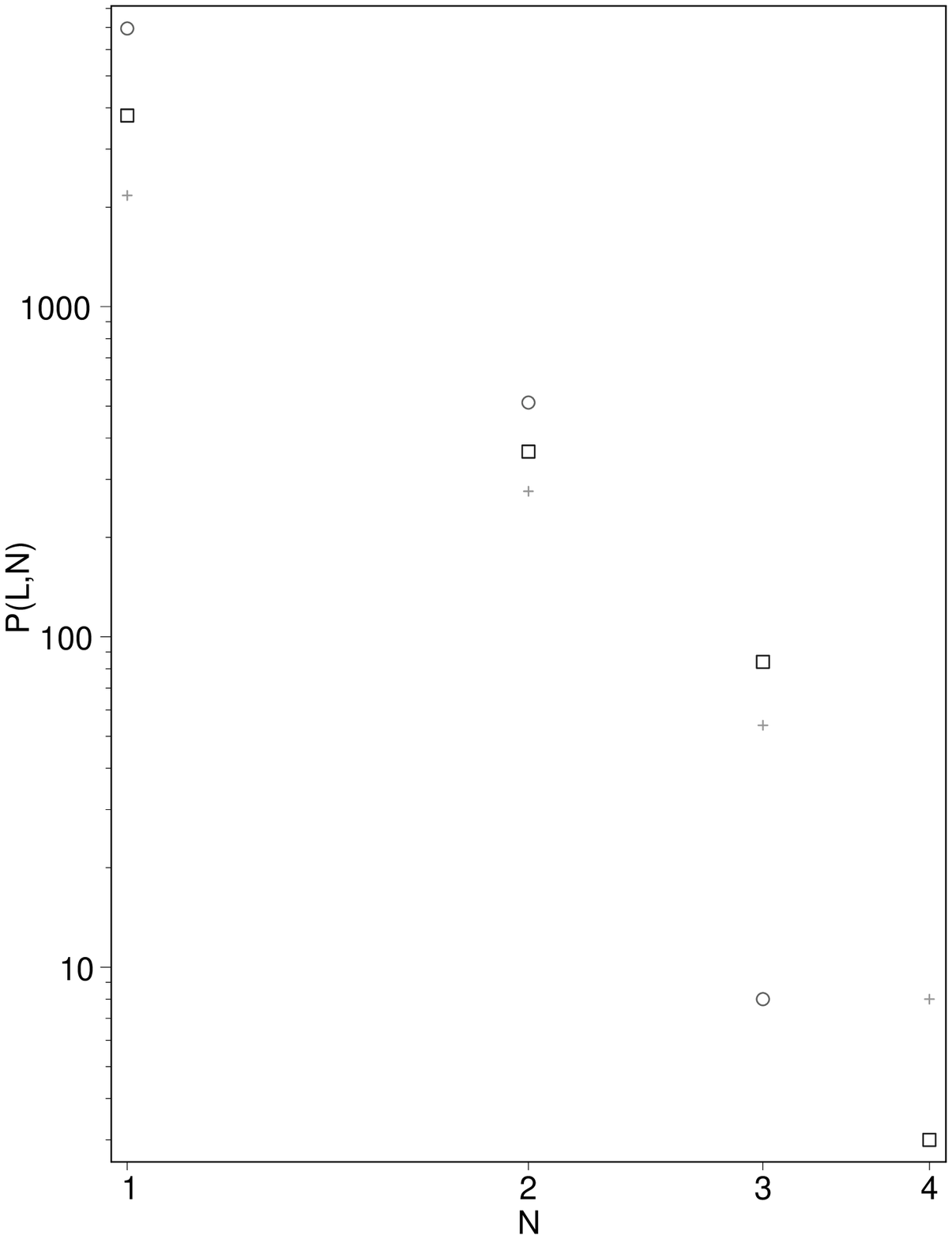}} \put(75,0)
{\parbox[t]{5mm}{\bf e.}} \put(215,10)
{\includegraphics[width=6cm,height=6cm]{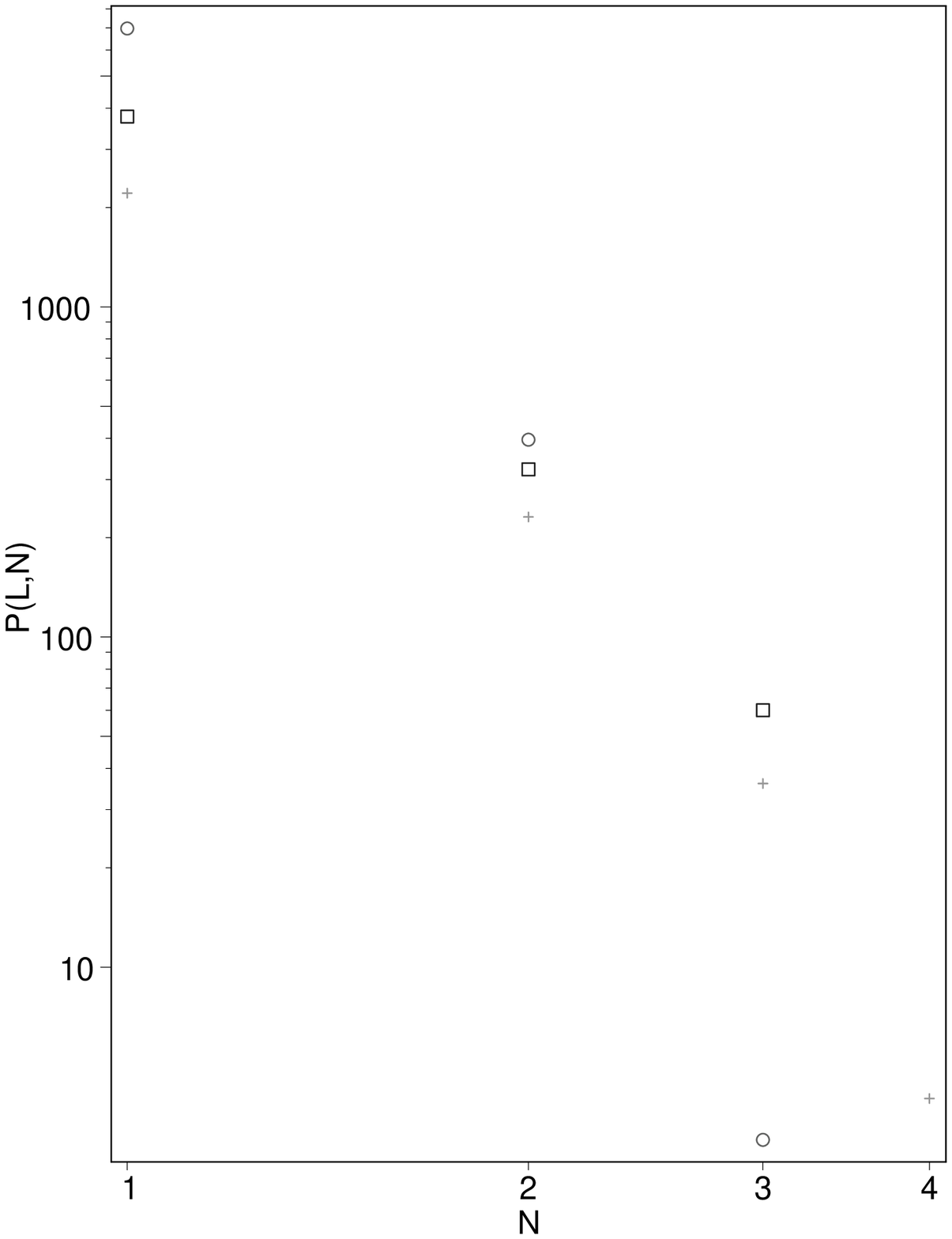}} \put(255,0)
{\parbox[t]{5mm}{\bf f.}}
\end{picture}
\caption{\label{fig4}
Number of sequences $P(L,N)$ of
length $L$ connected by $N$ lines.
{\bf a:} $L=1$,  {\bf b:} $L=2$, {\bf c:} $L=3$,
{\bf d:} $L=4$ {\bf e:} $L=5$,  {\bf f:} $L=6$.
Symbols as in the figure \protect\ref{fig2}.
}
\end{figure}

First, we analyse $P(1,N)$: the number of
segments of length 1 consisting of $N$ lines.
From now on we will calculate the integral
characteristics, that is calculating $P(1,N)$ we take
into account {\em all} sequences of stations of length 1
consisting of $N$ lines. The result is shown in figure \ref{fig4}.a.
Again, a power law holds:
\begin{equation} \label{2}
P(1,N) \sim N^{-\gamma_1},
\end{equation}
with the exponent $\gamma_1$ ranging from 3 to 4 for the networks considered.
The least squares fit gives the values given in table \ref{tab1}.
\begin{table}
\begin{center}
\begin{tabular}{|l|l|l|l|}
\hline
 $L$ & Berlin & D\"uss. & Paris \\
\hline 1 & 3,47 &   3,09 &   3,97 \\
 2 & 3,58 &   3,58 &   4,59 \\
 3 & 3,77 &   4,08 &   4,95 \\
 4 & 4,55 &   3,82 &   6,06 \\
 5 & 4,75 &   3,9  &   5,9  \\
\hline
\end{tabular}
\end{center}
\caption{\label{tab1} Scaling exponent $\gamma_L$ obtained by the
least-square fit for the number of sequences of length $L$
connected by $N$ lines $P(L,N)$ for the PT networks of Berlin,
D\"usseldorf and Paris. }
\end{table}
Obviously, the power-law behaviour found for $P(1,N)$ should not hold for
all values of $L$ and $N$. Indeed, the other limiting case $P(L,1)$ describes
distribution of lengths of different routes.
Function $P(L,1)$ will have a maximum corresponding to the mean length
of route, provided the network has enough routes for a good statistics.
The behaviour of the function $P(L,N)$ with increasing $L$ is shown  by
figures \ref{fig4}.
It is tempting to describe the plots given in the figures by power-laws
with exponents $\gamma_L$, depending on the line length $L$:
\begin{equation} \label{3}
P(L,N) \sim N^{-\gamma_L}.
\end{equation}
Numerical values of the corresponding exponents $\gamma_L$ are given
in the table \ref{tab1} for $L=1...5$. The columns of the table give
results obtained by the least-squares fit for each city separately.
The data of figures \ref{fig4}.a-f seem obey power laws, but
obviously it is too early to state this as a definite conclusion at
least for two reasons. The first obvious reason is that the statistics
considered so far concerns three different networks only (PT networks
of three different cities). Although the networks themselves were chosen
to be large enough (see section \ref{II}), still it is desirable to support
the data obtained by analysis of a larger number of networks. The second reason
is more subtle: it is obvious, that $P(L,N)$ decreases with $N$ for the
(real) PT networks considered here. Indeed, the larger the $N$ the
smaller $P(N)$ (\ref{1}) and the smaller the probability that
several ($L$) subsequent nodes have the same node degree. Therefore, the
number of data points for $P(L,N)$ always will decrease with $L$ and $N$
leading to poor statistics for any real network: c.f. number of data points in figures
\ref{fig4}.a and \ref{fig4}.f. So the data for $\gamma_L$
presented in the table \ref{tab1} for each separate network
 networks are to be considered rather as an attempt
for a power law fit, and not as a definite conclusion about the universal
power-law distribution. Nevertheless, we consider this observation to be
interesting for further study.

\section{Conclusions, outlook, and \dots best wishes! \label{V}}

This paper is written for a special issue of CMP in honour of Reinhard
Folk 60th birthday. The majority of the contributions to the Festschrift reflect
the honorees field of activity: phase transitions in condensed
matter physics. In our introduction we tried to show a possible link between
complex network behaviour and criticality in condensed matter provided by
the scaling phenomenon, as we, being new to the
fascinating field of complex networks understand it.
By this paper we want to congratulate
Reinhard Folk on the occasion of his birthday and to acknowledge
his vivid interest and active participation in the numerous discussions
about complex networks during the great time  two of us
(CvF and YuH) had enjoying his wonderful hospitality in
Wintersemester 2004 in Linz.

The PT networks discussed in this paper provide one more example
of the scale-free networks, as demonstrated by the power law
behaviour (\ref{1}) of their node degree distribution (figure
\ref{fig2}). Besides, some other properties of these networks may
obey scaling laws.  An example was given by the integrated
distribution of the neighbourhood size $Z_1$ describing the number
of stations which can be serviced from the chosen one without
making a change.  We found evidence that the functions $P(L,N)$ -
the number of node segments of length $L$ connected by $N$ lines -
may be described by power-law fits (\ref{3}) at least for low $L$.
Our interest in these values is caused by the fact, that they are
important to understand specific properties of PT networks, where
many routes possess common subsets of nodes and other examples of
similar structure that were briefly mentioned in the paper as
well.

A natural continuation of this study is to improve the network statistics
considered here, by taking a larger number of PT networks as well
as to continue analysis of different network characteristics.

One of us (YuH) acknowledges the Austrian Fonds zur F\"orderung der
wissen\-schaft\-li\-chen Forschung for support under project No. 16574-PHY.

\label{last@page}

\begin{thebibliography}{20}
\bibitem{books} For an introduction to networks which does not
postulate any professional preparation see e.g.:
 A.-L. Barab\'asi. Linked. Plume, 2003.
 A recent book which gives the essentials of
 the theory: S. N. Dorogovtsev, J. F. F. Mendes.
 Evolution of networks. Oxford University Press, Oxford, 2003
\bibitem{Barabasi02}
R. Albert, A.-L. Barab\'asi. Rev. Mod. Phys. {\bf 74} 47 (2002)
\bibitem{Barabasi99}
A.-L. Barab\'asi, R. Albert. Science {\bf 286} 509 (1999)
\bibitem{www}
R. Albert, H. Jeong, A.-L. Barab\'asi. Nature (London)
{\bf 401} 130 (1999)
\bibitem{cell}
H. Jeong, B. Tombor, R. Albert, Z.N. Oltvai, A.-L. Barab\'asi. Nature (London)
{\bf 407} 651 (2000)
\bibitem{social}
M.S.  Granovetter. Am. J. Sociol. {\bf 78} 1360 (1973)
\bibitem{ecol} J.M. Montoya, R.V. Sol\'e. preprint
cond-mat/0011195 (2000); R.V. Sol\'e, J.M. Montoya.
Proc. R.  Soc. Lond. B {\bf 268} 2039 (2001)
\bibitem{Zipf35}
G. K. Zipf. The psycho-biology of language. An introduction to
dynamic philology. Houghton Mifflin Comp., Boston, 1935.
\bibitem{critbooks}
C. Domb. The critical point. Taylor \& Francis, London, 1996; P.M.
Chaikin, T.C. Lubensky. Principles of condensed matter physics.
Cambridge University Press, Cambridge, 1995.
\bibitem{order}
Recent papers reviewing critical phenomena may be found in:
Yu. Holovatch (Ed.). Order, disorder and criticality. Advanced problems
of phase transition theory. World Scientific, Singapore, 2004.
\bibitem{rgbooks} J. Zinn-Justin. Quantum field theory and
critical phenomena. Oxford University Press, Oxford, 1996; H.
Kleinert, V. Schulte-Frohlinde. Critical properties of
$\phi^4$-theories. World Scientific, Singapore, 1991.
\bibitem{data} The data for the public transport of Berlin,
D\"usseldorf and other German cities may be fould following links
on http://www.oepnv.de/. Data for the Paris bus network used in our study
is available at http://www.citefutee.com/.
\bibitem{note1} Here, we take the node degree $k$ to be equal
to the number of PT routes going through the given node (which is
half the number of lines for the internal nodes crossed by these
routes).
\bibitem{Holyst04}
J.H. Holyst, J. Sienkiewicz, A. Fronczak, P. Fronczak, K. Suchecki.
preprint cond-mat/0411160, 2004.
\end{thebibliography}
\end{document}